\documentclass[final]{gGAF2e}
     \usepackage{datetime}
     
     \newtimeformat{isotime}{\twodigit{\THEHOUR}:\twodigit{\THEMINUTE}:\twodigit{\THESECOND}}
     \yyyymmdddate
     \settimeformat{isotime}
\usepackage[utf8x]{inputenc}
\usepackage{tikz}
\usepackage{epsfig}
\usepackage[normalem]{ulem}
\usepackage[colorlinks=true,breaklinks=true]{hyperref}
\usepackage{color}
\usepackage{graphicx}
\usepackage{natbib}
\usepackage{amsmath,amsfonts,amssymb}
\usepackage[mathlines]{lineno}
\bibpunct[; ]{(}{)}{,}{a}{}{;}
\usepackage{doi}
\usepackage{url}

\renewcommand{\Theta}{{\varTheta}}
\renewcommand{\Gamma}{{\varGamma}}
\renewcommand{\Xi}{{\varXi}}
\renewcommand{\Phi}{{\varPhi}}
\renewcommand{\Psi}{{\varPsi}}
\renewcommand{\Omega}{{\varOmega}}
\usepackage{cancel}
\allowdisplaybreaks
\reversemarginpar
\usepackage[percent]{overpic}

\usepackage{multiequations}
\newcommand*{\bse}{\begin{subequations}}
\newcommand*{\ese}{\end{subequations}}
\newcommand*{\bme}{\begin{multiequations}}
\newcommand*{\eme}{\end{multiequations}}
\newcommand*{\be}{\begin{equation}}
\newcommand*{\ee}{\end{equation}}

\renewcommand{\vec}[1]{\mbox{\boldmath $ #1$}}

\newcommand{\B}{\vec B}

\renewcommand{\r}{\vec r}
\renewcommand{\k}{\hat{\vec k}}

\newcommand{\pol}{v}
\newcommand{\tor}{w}

\newcommand{\Ra}{\mathrm{Ra}}
\newcommand{\RTi}{\mathrm{R}_T}

\newcommand{\RCi}{\mathrm{R}_C}

\renewcommand{\Pr}{\mathrm{Pr}}
\newcommand{\Sc}{\mathrm{Sc}}
\newcommand{\Pm}{\mathrm{Pm}}
\newcommand{\Rey}{\mathrm{Re}}
\newcommand{\Le}{\mathrm{Le}}

\providecommand{\Emp}{\overline{E}_p}
\providecommand{\Efp}{\widetilde{E}_p}
\providecommand{\Emt}{\overline{E}_t}
\providecommand{\Eft}{\widetilde{E}_t}

\usepackage{color}

\title{Regimes of thermo-compositional convection and related dynamos in rotating spherical shells}
\author
{James F.~Mather$^1$ and Radostin D.~Simitev$^2$
\thanks{$^1$ \href{https://orcid.org/0000-0001-9518-0515}{orcid.org/0000-0001-9518-0515}}
\thanks{$^2$
\href{mailto:Radostin.Simitev@glasgow.ac.uk}{Radostin.Simitev@glasgow.ac.uk},  \href{https://orcid.org/0000-0002-2207-5789}{orcid.org/0000-0002-2207-5789}} \\\vspace{6pt}
School of Mathematics and Statistics, University of Glasgow -- Glasgow
G12 8QQ, UK\\ 
\received{Submitted to GAFD 2019-07-16; Accepted to GAFD 2020-04-24} 
}

\begin{document}

\jvol{xx} \jnum{xx} \jyear{2020} \jmonth{}
\markboth{\rm J.F.~MATHER AND R.D.~SIMITEV}{\rm GEOPHYSICAL AND ASTROPHYSICAL FLUID DYNAMICS}

%\pagewiselinenumbers
%\linenumbers

\maketitle

\begin{abstract}
% Context
Convection and magnetic field generation in the Earth and planetary
interiors are driven by both thermal and compositional gradients.
% Methodology and factual
In this work numerical simulations of finite-amplitude double-diffusive
convection and dynamo action in rapidly rotating spherical shells full
of incompressible two-component electrically-conducting fluid are reported.
% Main Results and conclusions
% Convection
Four distinct regimes of rotating double-diffusive convection
identified in a recent linear analysis (Silva et al., 2019,
Geophys.~Astrophys.~Fluid Dyn., doi:10.1080/03091929.2019.1640875) 
are found to persist significantly beyond the
onset of instability while their regime transitions remain abrupt. 
In the semi-convecting and the fingering regimes characteristic flow
velocities are small compared to those in the thermally- and
compositionally-dominated overturning regimes, while zonal flows
remain weak in all regimes apart from the 
thermally-dominated one. Compositionally-dominated overturning
convection exhibits significantly narrower azimuthal structures
compared to all other regimes while differential rotation becomes the
dominant flow component in the thermally-dominated case as driving is
increased.
%Dynamos
Dynamo action occurs in all regimes apart from the regime of fingering
convection. While dynamos persist in the semi-convective regime they
are very much impaired by small flow intensities and very weak
differential rotation in this regime which makes poloidal to toroidal
field conversion problematic. The dynamos in the thermally-dominated
regime include oscillating dipolar, quadrupolar and multipolar cases
similar to the ones known from earlier parameter
studies. Dynamos in the compositionally-dominated regime exhibit
subdued temporal variation and remain predominantly dipolar due to
weak zonal flow in this regime.
% Importance
These results significantly enhance our understanding of the primary
drivers of planetary core flows and magnetic fields. 
\end{abstract}


\section{Introduction}

% topic
Convective flows and magnetic field generation in Earth's fluid outer
core are driven by a combination of thermal and chemical composition
gradients \citep{Kono2002,Jones2015,Wicht2019}. 
Rotating double-diffusive convection is also likely to occur in and
affect the magnetic properties of other planets including Mercury
\citep{Breuer2007}, Venus \citep{Jacobson2017}, Jupiter
\citep{Moll2017} and Saturn \citep{Leconte2012} as well as many
stellar objects \citep{Garaud2018}.  

% previous studies and lack of 
Numerical models of {global} planetary and core convection and dynamos have
been predominantly single-diffusive \citep{Jones2011}. Either purely
thermal convection was assumed or the so called ``co-density''
formulation {\citep{BraginskyRoberts1995} was used to effectively
replace temperature and concentration by a single field variable.
A notable exception is the celebrated early geodynamo model of
\citet{GLATZ&ROBS1996} where separate thermal and compositional
buoyancies were used but identical values for the thermal and chemical
diffusivity were set thus excluding double-diffusive effects}.  
However, single-diffusive convection models fail to account for
significant differences in the diffusivities of heat and chemical
constituents as well as for essential differences in boundary
conditions and sink/source distributions of the temperature and
of the constituent concentration field \citep{Jones2015}. 
In the last decade several authors have recognized these limitations
and have sought to investigate explicitly double-diffusive
{thermo}-compositional effects on convection flows and dynamo processes
in rapidly-rotating spheres and shells.  
Using two separate equations for the temperature and for the
concentration of light constituents, respectively, \citet{Breuer2010}
observed an abrupt change in convective regime when the relative
contribution of compositional driving exceeds 20\%.  
In a similar model, \citet{Trumper2012} considered the effect of
distinct boundary conditions for temperature and concentration and
obtained preliminary results on 
the onset of convection using an initial value code. 
Simulating a double-diffusive model of Mercury's dynamo, \citet{Manglik2010}
observed that when thermal and compositional buoyancy are of equal
intensity, a stratified outer layer is formed and is then penetrated by
fingering convection that in turn enhances the poloidal magnetic
field, a significant difference compared to co-density cases. 
Reporting some 20 numerical dynamo runs, \citet{Takahashi2014}
found {that, due to helicity increase,} magnetic fields have predominantly
non-dipolar morphology when compositional buoyancy is 
less than 40\% of the total driving and predominantly dipolar one
otherwise. In short, significant thermo-compositional effects were found
in all of the latter works. However, these studies were largely limited 
to numerical runs isolated in the configuration space and considered
the case of when the thermal buoyancy and the compositional buoyancy
are both destabilising.  

% need for systematic studies
Systematic parameter studies  of the linear onset of double-diffusive convection
in rotating spherical shells were undertaken by \cite{Net2012} and
recently by \citet{Silva2019}.
% our own findings that we wish to pursue further
Following \citet{Simitev2011}, the study of \citet{Silva2019} confirmed
that due to distinct ``double-diffusive'' eigenmodes, critical curves for the
onset of thermo-compositional instability are generally multi-valued and form what may be
described as ``pockets'' of instability protruding regions of
quiescence. Situations thus arise whereby increasing the thermal or
the compositional driving leads to onset of instability at the entry of a
pocket and a subsequent return to quiescence as driving is further
increased to exit the pocket.
The pockets of instability are closely related to transitions between
four different regimes of rapidly-rotating double-diffusive convection
that can be identified 
as semi-convection, fingering convection, thermally-dominated overturning
convection and chemically dominated overturning convection. The onset
of these regimes were mapped by \citet{Silva2019} who probed a
significant portion of the configuration space by varying the values
of all governing parameters.  

% aims and goals
{
The main goals of the present work are (a) to establish the possibility of dynamo
action arising due to double-diffusive flows on stably stratified
background, in other words in the regimes of semi-diffusive and
fingering flows, and (b) to assess how dynamos generated by thermally-
and chemically-dominated overturning convection differ from each
other.
Before these goals can be addressed, it is necessary to (c)
trace out the flow regime boundaries in the non-linear domain.
While rotating double-diffusive convection at finite amplitude was recently
studied by \citet{Monville2019}, essential differences with our
model remain since full spherical geometry, no-slip velocity
conditions and mainly the semi-convective regime were considered by
these authors. 
Thus, using as a starting point the linear results of \cite{Silva2019},}
we perform parameter continuation increasing the thermal and 
compositional Rayleigh numbers (defined further below) beyond
onset. For the sake of comparison, we keep most other governing
parameters fixed to values where purely thermal convection and dynamo solutions
are well studied in the literature. Even with these restrictions,
results from over 80 new finite-amplitude thermo-compositional
convection simulations and over 30 dynamo simulations are summarized below.
We take the opportunity, to assess to what extent instability pockets
survive nonlinear interactions; the structure and intensity of zonal
and other flow components; observe how magnetic field morphology and
symmetry change across the regime boundaries and describe the
time-dependent behaviour of solutions and investigate whether dynamos
close to each other in the parameter space can exhibit widely
different morphology and behaviour. These questions are relevant to
understanding the geomagnetic field at present and in geological time
as well as to understanding the zoo of other planetary and stellar
magnetic fields. 

{Beyond spherical models, there has been a significant interest in astrophysical applications
of double-diffusive convection in recent years \citep{Garaud2018}. 
Asymptotic scaling laws for turbulent heat and compositional transport
have been proposed and spontaneous layering has been investigated at
small values of the Prandtl number for both the fingering regime
\citep{Traxler2011,Brown2013} and the semi-convective regime
\citep{Mirouh2012,Wood2013,Moll2016}. These results complement our
analysis but direct application to the case reported here remains
difficult. This is because  a local ``unbounded gradient layer'' model \citep{Radko2013} in a
triply periodic box is used in the latter works to minimize the effect
of boundaries and to exclude overturning flows from the analysis. In
contrast, we aim to capture all regimes of global-scale rotating
thermo-compositional convection in spherical geometry and thus employ
a ``vertically bounded layer'' model of the type proposed by
\cite{Veronis1968}.}

\section{Mathematical model}
We follow standard mathematical formulations of the problem of
rotating  spherical dynamos e.g.~\citep{Simitev2005a} modified by introducing a
separate equation for the chemical concentration of the form 
used in \citep{Silva2019}.
In detail, we consider a two-component electrically conducting fluid confined to
a spherical shell rotating with a fixed angular velocity $\Omega \k$,
where $\k$ is the unit vector in the direction of the axis of rotation. 
The inner and outer spherical surfaces, $r=r_i$ and $r=r_o$,
respectively, are kept at constant values of the temperature
and of the concentration of the light element.
The gravity field is assumed in the form $\vec g = - \gamma d \r$
where $\r$ is the position vector with respect to the center of the
sphere and $r$ is its length measured in units of the thickness $d$ of
the spherical shell.
Assuming volumetric sources/sinks of thermal and compositional buoyancy 
with constant densities $\beta_T$ and $\beta_C$, respectively, a
static state exists with temperature and concentration profiles given
by 
\begin{align}
T_S=T_0-\dfrac{\beta_T}{2}r^2,  \enspace\enspace
C_S={C_0}-\dfrac{\beta_C}{2}r^2,
\label{eq:backgroundTC}
\end{align}
respectively. Here $T_0$ and $C_0$ are constant reference values of
temperature and concentration.
We employ the Boussinesq approximation in that all material properties
of the fluid are assumed constant except the density which is taken to 
depend on temperature and concentration so the following truncated
Taylor expansion near its reference value $\rho_0$ is used when it
enters the buoyancy term 
$$
\rho = \rho_0 (1- \alpha_T \Theta - \alpha_C \Gamma), 
$$
where $\alpha_T$ and $\alpha_C$ are the specific
thermal and compositional coefficients of expansion/contraction, and $\Theta$ and
$\Gamma$ are the deviations from the temperature and the concentration
basic static states $T_S$ and $C_S$ given by equations
\eqref{eq:backgroundTC}.
The length $d$, the time $d^2 / \nu$, the 
temperature $\nu^2 / \gamma \alpha_T d^4$ and composition  $\nu^2 / \gamma \alpha_C d^4$ are used as scales for the
dimensionless description of the problem where $\nu$ denotes the
kinematic viscosity.
The dimensionless governing equations of momentum, temperature,
concentration, magnetic induction and the conditions of
incompressibility of the fluid and solenoidality of the magnetic field
are then given by 
\begin{subequations}
\label{equations1}
\begin{gather}
\label{1a}
(\partial_t + \vec u \cdot {\bm{ \nabla}}) \vec u + \tau \vec k \times
\vec u = - {\bm{ \nabla}} \pi + (\Theta+ \Gamma) \vec r +  \nabla^2 \vec u + ({\bm{ \nabla}} \times \vec B) \times \vec B, \\
\label{1b}
{\bm{ \nabla}} \cdot \vec u = 0,\\
\label{1cT}
\Pr(\partial_t + \vec u \cdot {\bm{ \nabla}}) \Theta = \RTi \vec r \cdot \vec u + \nabla^2 \Theta,\\
\label{1cC}
\Sc(\partial_t + \vec u \cdot {\bm{ \nabla}}) \Gamma = \RCi \vec r \cdot \vec u + \nabla^2 \Gamma, \\
 \label{1dmag}
\partial_t \vec B =  {\bm{ \nabla}} \times  \left(\vec u \times
    \vec B \right) + \Pm^{-1} \nabla^2 \vec B,\\
\label{1dsol}
{\bm{ \nabla}} \cdot \vec B = 0,
\end{gather}
\end{subequations}
where $\vec u$ and $\vec B$ are the velocity and
the magnetic field vectors, respectively, and $\pi$ denotes an effective pressure field
representing all terms that can be expressed as a gradient. Eight dimensionless numbers appear in
the equations, namely, the shell radius ratio, the thermal
Rayleigh number $\RTi$,  the compositional Rayleigh number $\RCi$, the
Coriolis parameter $\tau$, the Prandtl number $\Pr$, the Schmidt number $\Sc$
and the magnetic Prandtl number $\Pm$ defined as
\begin{gather}
{\eta=\frac{r_i}{r_o}}, \enspace
\RTi = \frac{\alpha_T \gamma \beta_T d^6}{\nu \kappa} ,  \enspace  
 \RCi = \frac{\alpha_C \gamma \beta_C d^6}{\nu D}, \enspace 
\label{1eother}
 \tau =
\frac{2 \Omega d^2}{\nu} , \enspace \Pr = \frac{\nu}{\kappa}, \enspace \Sc = \frac{\nu}{D}, \enspace \Pm=\dfrac{\nu}{\lambda},
\end{gather}
respectively. Here, $\kappa$ denotes the thermal diffusivity, $D$
denotes the mass diffusivity and $\lambda$ denotes the magnetic
diffusivity. 

Since the velocity and the magnetic field are both solenoidal the general representation
in terms of poloidal and toroidal components can be used, 
\begin{displaymath}
\vec u = {\bm{ \nabla}} \times ( {\bm{ \nabla}} v \times \vec r) + {\bm{ \nabla}} w \times
\vec r, \enspace \enspace 
\vec B = {\bm{ \nabla}} \times ( {\bm{ \nabla}} h \times \vec r) + {\bm{ \nabla}} g \times
\vec r.
\end{displaymath}
By multiplying the (curl)$^2$ and the curl of the momentum equation
(\ref{1a}) by $\vec r$ we obtain two equations for the poloidal and
toroidal scalar fields of the velocity, $v$ and $w$,
\begin{subequations}
\label{poltoreq}
\begin{gather}
\label{2a}
\big(( \nabla^2 - \partial_t) L_2 + \tau \partial_{\varphi} \big) \nabla^2 v +
\tau Q w - L_2 (\Theta +\Gamma )=  - \vec r \cdot {\bm{ \nabla}} \times  {\bm{ \nabla}} \times
( \vec u \cdot {\bm{ \nabla}} \vec u -{\bf B}\cdot{\bm{ \nabla}}{\bf B}), \\
\label{2b}
\big((\nabla^2 - \partial_t) L_2 + \tau \partial_{\varphi} \big) w - \tau Qv
= \vec r \cdot {\bm{ \nabla}} \times ( \vec u \cdot {\bm{ \nabla}} \vec u-{\bf B}\cdot{\bm{ \nabla}}{\bf B}).
\end{gather}
The temperature and the concentration equations may be written as follows
\begin{gather}
\label{2c}
\nabla^2\Theta + \RTi L_2v=\Pr(\partial_t+{\bf u}\cdot{\bm{ \nabla}})\Theta,\\
\label{2d}
\nabla^2\Gamma + \RCi L_2v=\Sc(\partial_t+{\bf u}\cdot{\bm{ \nabla}})\Gamma.
\end{gather}
The equations for ploidal and toroidal scalar of the magnetic field,
$h$ and $g$, are obtained by multiplication of \eqref{1dmag}
and of its curl by $\bf{r}$ 
\begin{gather}
\label{2e}
\nabla^2L_2h=\Pm\big(\partial_tL_2h-{\bf r}\cdot {\bm{ \nabla}} \times({\bf u}\times{\bf B})\big),\\
\label{2f}
\nabla^2L_2g=\Pm\big(\partial_tL_2g-{\bf r}\cdot {\bm{ \nabla}} \times{\bm{ \nabla}} \times({\bf u}\times{\bf B})\big).
\end{gather}
\end{subequations}
In the above, $\partial_t$ and $\partial_{\varphi}$ denote the partial
derivatives with respect to time $t$ and with respect to the  angle
$\varphi$ of a spherical system of coordinates $r, \theta, \varphi$
and the operators $L_2$ and $Q$ are defined by  
\begin{gather*}
\label{3a}
L_2 \equiv - r^2 \nabla^2 + \partial_r (r^2 \partial_r),\\
\label{3b}
Q \equiv r \cos \theta \nabla^2 - (L_2 + r \partial_r ) ( \cos \theta
\partial_r - r^{-1} \sin \theta \partial_{\theta}). 
\end{gather*}

Stress-free boundaries with fixed temperature and concentration values
are assumed,
\begin{subequations}
\label{bceq}
\begin{equation}
\label{4}
v = \partial^2_{rr}v = \partial_r (w/r) = \Theta =\Gamma= 0 
\quad 
\mbox{ at } \enspace r=r_i, \enspace r=r_o,
\end{equation}
where $r_i = \eta / (1- \eta)$ and $r_o = (1-\eta)^{-1}$.
For the magnetic field, electrically insulating boundaries are assumed
in such a way that the poloidal function $h$ is matched to the
function $h^{(e)}$ which describes the potential fields outside the
fluid shell 
\begin{align}
\label{5}
g=h-h^{(e)}=\partial_r(h-h^{(e)})=0 \quad \text{ at }r=r_i,\enspace r=r_o.
\end{align}
\end{subequations}
{The use of Dirichlet conditions for temperature and concentration
and of stress-free boundaries is a modeling simplification made for 
several reasons. First as discussed, we essentially rely on the linear
results of \citet{Silva2019} which were obtained with this choice of
boundary conditions in order to isolate the effects induced by 
differences in thermal and concentration diffusivities from the
effects induced by differences in boundary conditions and in source/sink
distributions. Second, there is no universal agreement on the most
appropriate choice of boundary conditions even for the best-known
planetary core, that of the Earth. In Earth’s core, rigid flow
boundary conditions at both boundaries seem appropriate. However, in common with all
simulations reported in the literature, the Coriolis number $\tau$ we
use is many orders of magnitude too small  and the viscous Ekman
boundary layers, whose thickness scales like $\tau^{-1/2}$ becomes
much more dominant in the simulations than it is in reality. 
We  follow e.g.~\citep{Kuang1997,Simitev2011b,Aubert2012} and use
stress-free conditions to alleviate this. A detailed comparison 
of no-slip and stress-free models of the dynamics of rapidly rotating
fluid in a spherical shell has been reported by \citep{Livermore2016},
also see the review of \citet{Roberts2013}.
The inner core is a source for heat and light elements and it is
appropriate to model this by fixed flux conditions for $\Theta$ and
$\Gamma$. Indeed, the composition flux has to vanish at the
core-mantle boundary as chemical elements cannot penetrate it. Heat
flux at the core-mantle boundary can be related to lower mantle
temperature as the mantle evolves on timescales much longer
than these of the outer core. However, the thermal and compositional
boundary conditions at the interface to a growing inner iron core are
likely even more involved, in that, the temperature flux and the
compositional flux are related to each other via a coupled set of
differential equations
\citep{BraginskyRoberts1995,GLATZ&ROBS1996}. Furthermore, it is
uncertain how a seismologically distinct mushy or slurry F-layer at base of Earth’s
outer core may be \citep{Wong2018} affect boundary conditions. The
outer boundary of the convective regions of gas planets and stars
requires yet different boundary conditions, see discussion in
\citep{Wicht2010}. Various choices of thermal and compositional
boundary conditions were investigated by \citep{Hori2012}.}

\section{Numerical methods and diagnostics}

\subsection{{Methods and code}}
For the direct numerical integration of the problem defined by the scalar
equations \eqref{poltoreq} and the boundary conditions \eqref{bceq} we use
a pseudo-spectral method described by \cite{Tilgner1999}. A code
used by us for a number of
years \citep{Busse2003a,Simitev2011b,Simitev2015} and benchmarked for
accuracy most recently in \citep{Marti2014,Matsui2016}  was adapted to include  
the additional equation for the chemical concentration of light
elements \eqref{2d}. The code has been made open source \citep{silva2018b}.
We briefly mention here that the spatial discretisation is based on an
expansion of all dependent variables in spherical harmonics for
the angular dependences and in Chebychev polynomials for the radial
dependence, e.g. the expansion of the poloidal scalar function takes
the form 
\begin{equation}
\label{6}
v(r,\theta, \varphi) = \sum \limits_{l,m,n}^{N_l,N_m,N_n} V_{l,n}^m
(t) T_n\big(x(r)\big) {\mathrm P}_l^m (\theta) \exp ({\mathrm i} m \varphi),
\end{equation}
with $x(r) = 2(r-r_i)-1$, and analogous expressions for the other
dynamical variables, $w$, $h$, $g$, $\Theta$ and $\Gamma$ are
used. Here ${\mathrm P}_l^m$ denotes
the associated Legendre polynomials of degree $l$ and order $m$ and
$T_n$ denotes the Chebychev 
polynomials of degree $n$. The computation of nonlinear terms in spectral space is,
however, expensive so all nonlinear products and the Coriolis term are
computed in the physical space and then projected onto the
spectral space at every time step. 
A hybrid combination of the Crank--Nicolson scheme for the diffusion
terms and the second order Adams--Bashforth scheme for the nonlinear
terms is used for time-stepping.

\subsection{{Resolution and convergence}}
Spherical harmonics truncation of $N_l=N_m=144$ and up to $N_r=51$
collocation points in radial direction have been used and a time step
of the order $10^{-6}$ was typically required for the computations
reported below.
{
Calculations are considered resolved when the spectral
power of kinetic and magnetic energy drops by more than 
a factor of 100 from the spectral maximum to the cut-off wavelength, a
criterion commonly used e.g.~\citep{CHRIST&OLSON&GLATZ1999}. Typical
spectra are shown further below in figure \ref{fig08}. We have
evolved dynamo simulations for at least one ohmic diffusion time and 
hydrodynamic simulations for at least one viscous diffusion
time in most cases for much longer as seen in the time series included in
figure \ref{fig08}. All time averages are calculated after removing
transient periods of the simulations.
}

\subsection{{Diagnostics}}
To analyse the properties of the solution we typically monitor kinetic
and magnetic energy components. In particular, we decompose the
kinetic energy density into poloidal and toroidal parts
(respectively denoted by subscripts as in $X_p$ and 
$X_t$ in equations \eqref{Engs} below and where $X$ denotes an
appropriate quantity)  and further into mean (axisymmetric) and
fluctuating (nonaxisymmetric) components {(respectively denoted by
  bars and tildes as in $\overline X$ and $\widetilde X$ below)}  
and into equatorially-symmetric and equatorially-antisymmetric
components {(respectively denoted by superscripts as in $X^s$ and $X^a$ below)}, 
\begin{align}
\Emp = \overline E_p^s +\overline E_p^a &= {\textstyle\frac{1}{2}}\displaystyle
\left\langle \big({\bm{ \nabla}} \times ( {\bm{ \nabla}} {(\overline\pol^s+\overline\pol^a)} \times \vec r)
\big)^2  \right\rangle, \nonumber\\
\label{Engs}
\Emt = \overline E_t^s +\overline E_t^a& =
     {\textstyle\frac{1}{2}}\displaystyle
\left\langle \big({\bm{\nabla}} r {(\overline\tor^s+\overline\tor^a)} \times \vec r \big)^2
  \right\rangle,  \\
\Efp = \widetilde E_p^s +\widetilde E_p^a&=
     {\textstyle\frac{1}{2}}\displaystyle
\left\langle
\big({\bm{ \nabla}} \times ( {\bm{ \nabla}} {(\widetilde
  \pol^s+\widetilde \pol^a)} \times \vec r)  \big)^2 \right\rangle, \nonumber\\
\Eft = \widetilde E_t^s +\widetilde E_t^a&=
     {\textstyle\frac{1}{2}}\displaystyle
\left\langle \big(
           {\bm{ \nabla}} r {(\widetilde \tor^s+\widetilde
             \tor^a)} \times\vec r \big)^2  \right\rangle, \nonumber
\end{align}
where  angular brackets $\langle\,\, \rangle$ denote averages 
over the volume of the spherical shell. Since in our code the spectral 
representation of all fields $X$ is given by the set of coefficients
$\{X_l^m\}$ of their expansions in spherical harmonics $Y_l^m$, it is
easy to extract the relevant components, i.e. coefficients 
with $m=0$ and with $m\ne0$ represent axisymmetric and nonaxisymmetric
components, respectively, while coefficients with even $(l+m)$ and with
odd $(l+m)$ represent equatorially-symmetric and
equatorially-antisymmetric components, respectively. The magnetic
energy density is similarly decomposed into components.
{The time averaged Reynolds number is defined as
$$
\Rey=\sqrt{2 E},
$$
where $E=\Emp+\Emt+\Efp+\Eft$, is the total kinetic energy density.}
\begin{figure*}
\includegraphics[width=\textwidth]{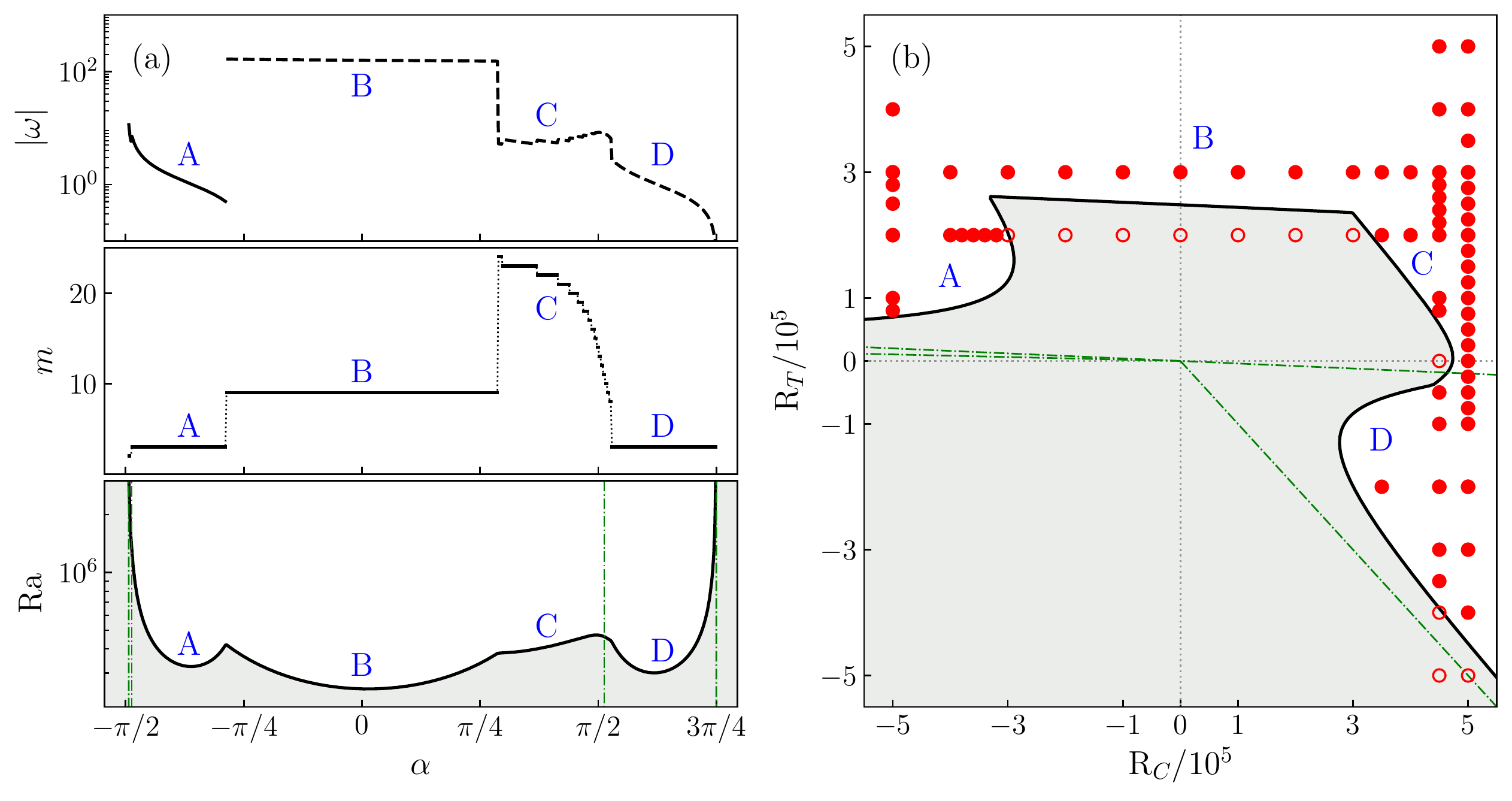}
\caption{Critical curves for the onset of convection and selected
finite-amplitude non-magnetic runs in the case $\Pr=1$, $\Sc=25$, $\tau=10^4$,
$\eta=0.35$.
(a) Critical curves as a function of the mixing angle $\alpha$ as
follows. Top panel: the magnitude of the critical drift rate
$|\omega|$ at onset; prograde direction (negative drift rate) indicated
by broken line, retrograde direction (positive drift rate)
indicated by solid line. 
Middle panel: values of the critical azimuthal wave-number $m$ at
onset. Bottom panel: values of the critical effective Rayleigh number
$\Ra$. 
(b) The critical curve in the $\RCi - \RTi$ plane is shown as a black
solid line. The positions in the parameter space of selected
supercritical and subcritical finite-amplitude runs are indicated with
full and empty red circles, respectively.
In both panels (a) and (b): The subcritical region of no convection is
shaded in light-grey. Blue letters A, B, C, and D denote 
the approximate location of the regimes of rotating semi-convection,
rotating thermally-dominated overturning convection, rotating
compositionally-dominated overturning convection and rotating fingering convection,
respectively, as discussed in the text. {In clockwise direction in (b)
and from left to right in the bottom panel of (a) green dash-dotted
lines mark the stability bounds 
$\RTi=-(\Pr+1/\Le)\RCi/(\Le(\Pr+1))$, 
$\RTi=-\RCi/\Le$,
$\RTi=-\RCi/\Le$ and
$\RTi=-\RCi$, respectively, see equations \eqref{bounds}.} 
(Colour online)
}
\label{fig1}
\end{figure*}

\section{The onset of rotating double-diffusive convection}

%Past results
A systematic investigation of the onset of double-diffusive convection
in rotating spherical shells is reported in \citep{Silva2019}. 
In this section, we summarize very briefly selected linear results 
of this study {since these will be used as a base for tracing out the
boundaries of thermo-compositional regimes in the non-linear
domain.} 

\subsection{{Onset at selected parameters}}
Figure \ref{fig1} shows the critical Rayleigh numbers $\RTi$ and
$\RCi$, the preferred azimuthal wave number $m$ and drift rate $\omega$ 
for the onset of instability at other parameter values fixed to 
\begin{gather}
\Pr=1, \enspace \Sc=25, \enspace \tau=10^4, \enspace
\eta=0.35, \enspace \Pm=0.
\label{fixedparms}
\end{gather}
{The Lewis number defined as $\Le=\Sc/\Pr$ is sometimes used in the
literature instead of the Schmidt number $\Sc$,
e.g. \citep{Simitev2011,Monville2019}. In our study the
value of the Lewis number is kept fixed to $\Le=25$ throughout.}
To facilitate direct comparison with \citep{Silva2019}, in the left
panel (a) of the figure a re-parametrisation is used where a `total
effective' Rayleigh  number and a mixing angle $\alpha$ are introduced
as follows 
$$
\Ra=\sqrt{\RTi^2+\RCi^2}, \qquad \alpha=\text{atan2}(\RCi,\RTi).
$$
Here the function $\text{atan2}(y,x)$ is defined for $x\in\mathbb{R}$,
$y\in \mathbb{R}$ as the principal argument $\text{Arg}(z)$ of the
complex
number $z=x+iy$, a notation used in many programming languages.
The more conventional representation of the critical curves  in the
$\RCi - \RTi$ plane is shown in the right panel figure \ref{fig1}(b).
Four convective regimes are immediately discernible 
-- semi-convection, thermally-dominated overturning
convection, compositionally-dominated overturning convection and
fingering convection. For convenience, these regimes are labeled by A,
B, C and D, respectively in figures and discussion throughout.
Regime A occurs for values of the mixing angle $\alpha$ between
$-\pi/2$ and $-5\pi/16$. This regime was identified in \citep{Silva2019} as semi-convection as
it occurs at negative values of $\RCi$ and positive values of $\RTi$
which in our model corresponds to cooler and lighter fluid over warmer
and heavier fluid. The linear analysis indicates that flows in this
regime consist of large spatial structures with azimuthal wave number
$m=3$ drifting in retrograde direction $(\omega>0)$. 
An abrupt transition to a different regime B occurs at values of the mixing angle of about $\alpha\approx
-5\pi/16$ and persists up to $\alpha\approx 5\pi/16$. Within this range
the critical curve is smooth and includes as a particular case purely thermal convection
$\RCi=0$ at $\alpha=0$ so this regime can be identified as
thermally-dominated overturning convection. At values of the
other governing parameters given by \eqref{fixedparms}  the flow takes
the familiar form of columnar convection with $m=9$ vertical columns
arranged in a cartridge belt inside the cylinder  tangent to the inner
core and drifting in prograde direction 
$(\omega<0)$. 
A second abrupt transition takes place at $\alpha\approx 5\pi/16$ and
brings the flow to a regime C that persist up to about $\alpha =
17\pi/32$. This range includes as a particular case purely
compositional convection $\RTi=0$ at 
$\alpha=\pi/2$ and is, therefore, identified as
compositionally-dominated overturning convection. It is interesting
to note that there is a large jump in the azimuthal wave number to
$m=25$ which then monotonically decreases to $m=8$ as the mixing angle
$\alpha$ is increased. The convective patterns of regime C continue to
drift in prograde direction.  
At $\alpha\approx 17\pi/32$ a last transition is observed leading to
regime D that persist to $\alpha = 3\pi/4$. This regime is identified
as fingering  convection as it largely occupies a region where
$\RTi<0$ and $\RCi>0$ corresponding to a configuration of warmer
and heavier fluid over cooler and lighter fluid. The wavenumber
assumes a constant value of $m=3$ and the flow pattern drifts in
prograde direction. 
We note that while the critical curve is single-valued when plotted in the
$\Ra-\alpha$ plane, it is multivalued in certain regions when plotted
in the $\RCi-\RTi$ plane. 
The linear results outlined above are robust in the sense that they
persist in a similar form in a significant range of parameter
values. For instance, \citet{Silva2019} report the same qualitative
picture for values of the Coriolis parameter in the range
$\tau\in[10^3,10^6]$ (becoming more pronounced for larger values of
$\tau$), for values of the Prandtl number in the range $\Pr \in
[10^{-5},10^{3}]$ (most pronounced at intermediate values of $\Pr$),
for Schmidt to Prandtl number ratio at least within the range $\Sc/Pr
\in[25,100]$, and for shell radius ratio in the range $\eta \in
[0.1,0.7]$.

\subsection{{Comparison with ideal estimates}}
{
It is of interest to compare these results with bounds for convective
and double-diffusive instabilities known for perferct
fluids. These can be summarized conveniently using the so called
stability density ratio, $R_0 := (\alpha_T |\nabla T_S|) /(\alpha_C
|\nabla C_S|)$, 
\citep{Stern1960}. Overturning convection, 
semi-convection and fingering  instabilities occur for values of $R_0$
in the intervals 
\begin{gather}
\label{bounds}
R_0<1, \qquad R_0 \in
\left(\frac{\Pr+\Le^{-1}}{\Pr+1}, 1\right), \qquad R_0 \in (1, \Le), 
\end{gather}
respectively, see e.g.~\citep{Radko2013,Garaud2018},
where in terms of our non-dimensional parameters the density ratio can
be expressed as $R_0 = \Le|\RTi/\RCi|$ at $r=1$.
The bound for the onset of overturning convection ($R_0 < 1$) derives
from the \cite{Ledoux1947} criterion for unstable stratification,
$N^2=N_T^2+N_C^2<0$ where $N_T^2 := -\alpha_T
\gamma \r\cdot\nabla T_S$ is the thermal contribution to the
Brunt-V\"ais\"al\"a frequency $N$, and $N_T^2 := -\alpha_C \gamma
\r\cdot\nabla C_S$, is the compositional 
contribution, respectively. Bounds \eqref{bounds} are shown in
figures \ref{fig1} and \ref{fig010}. Our results indicate that
fingering instability occurs in stably-stratified fluids $(R_0 > 1)$
in a region well captured by the bounds $R_0 \in (1,\Le)$.
In contrast, the semi-convection instability becomes dominant soon
after the $|\RTi|=|\RCi|$ line is crossed and well inside the 
unstably-stratified region $(R_0 < 1)$. This behaviour is due to the
fact that the additional buoyancy provided by the compositional
gradient counteracts to some extent the convection-inhibiting Coriolis
force as shown by \cite{Busse2002} and \cite{Simitev2011}.
Bounds \eqref{bounds} must, of course, be regarded as approximate
only. The Ledoux criterion is derived on the assumptions that the
diffusivities and viscosity are neglected, and bounds \eqref{bounds}
do not take into account rotation and spherical geometry effects,
all essential in our configuration. 
}

\section{Finite amplitude thermo-compositional convection}

{In order to investigate dynamo action in the four
distinct regimes of thermo-compositional convection it is first
necessary to determine the regions in the parameter space that these occupy
at finite-amplitides. To trace all regimes systematically while
keeping} the volume of simulations at bay, we fix most of the
parameter values as specified in equation \eqref{fixedparms} and  
vary the values of the thermal and compositional Rayleigh numbers
along several selected ``slices'' through the $\RCi - \RTi$ plane as
is best illustrated in figure  \ref{fig1}(b). 

\begin{figure*}
\centering
{\includegraphics[width=\textwidth]{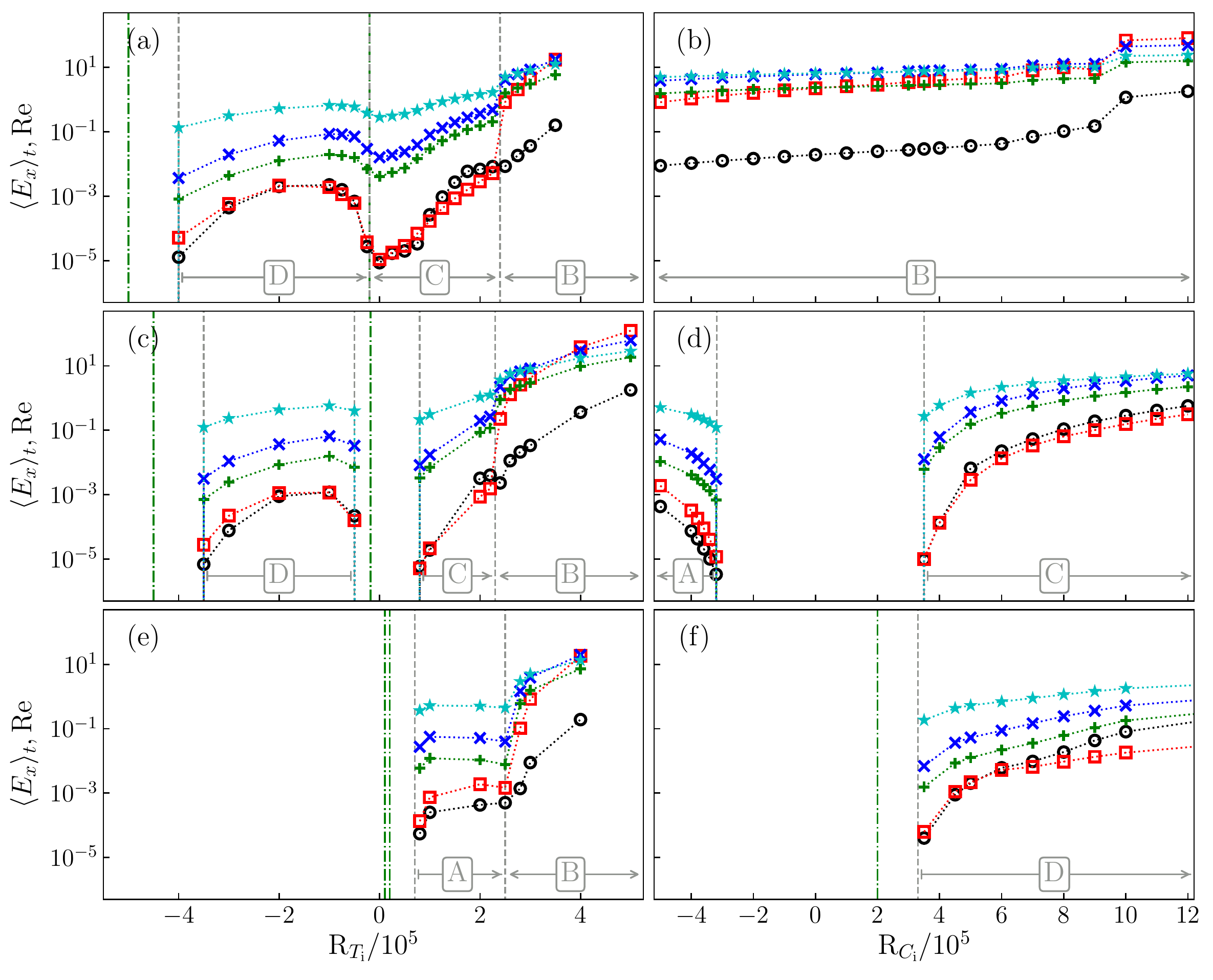}}
\caption{Time-averaged {Reynolds number and} kinetic energy density components in the case
$\Pr=1$,  $\Sc=25$,  $\tau=10^4$, $\eta=0.35$ and
(a) $\RCi=5\times10^5$, (b) $\RTi=3\times10^5$, (c) $\RCi=4.5\times10^5$, (d)
$\RTi=2\times10^5$, (e) $\RCi=-5\times10^5$, (f) $\RTi=-2\times10^5$ as functions of
$\RTi$ or $\RCi$ as appropriate. {The time-averaged Reynolds numbers
are denoted by cyan star symbols.} The time-averaged mean poloidal
component $\langle \Emp\rangle_t$,
mean toroidal component $\langle\Emt\rangle_t$, 
fluctuating poloidal component $\langle\Eft\rangle_t$ and fluctuating toroidal
component $\langle\Eft\rangle_t$ are indicated  by a black circles, red squares, green
plus signs and blue cross signs, respectively.
Vertical broken lines indicate points of transition between
convective regimes. The ranges over which distinct regimes are
observed are indicated by arrows near the bottom abscissa. Arrowheads
indicate that convection extends further and ticks at end of arrows indicate
onset of convection. The locations in the $\RCi-\RTi$ plane of some
data points used here are also denoted by red solid circles in figure \ref{fig1}.
{Where they are in the frame of the plots, stability bounds \eqref{bounds} are
  marked by vertical green dash-dotted lines from left to right as
  follows (a,c) $\RTi=-\RCi$ and $\RTi=-\RCi/\Le$, (e)
$\RTi=-(\Pr+1/\Le)\RCi/(\Le(\Pr+1))$ and $\RTi=-\RCi/\Le$, (f) $\RCi=-\RTi$.}
(Colour online)
}
\label{fig010}
\end{figure*}

{Growth rates near the onset of convection will not be
  reported. This is because on one hand, the linear eigensolver method 
of \cite{Silva2019} is configured to compute the marginal stability
curve where, by definition, the growth rate $\sigma=0$. This is done by performing a
non-trivial numerical extremisation and parameter continuation
procedure as discussed in relation to figures 1 to 3 of
\citep{Silva2019}. On the other hand, most of our nonlinear simulations
use as their initial condition the nearest available equilibrated
nonlinear solution in order to avoid long transient periods in this
already computationally-expensive problem. However, nonlinear
simulations are in excellent agreement with the values of the critical
Rayleigh numbers predicted by the linear analysis as seen in both figures
\ref{fig1}  and \ref{fig010}.}

\subsection{{Regime boundaries}}
Earlier studies of purely thermal convection
e.g.~\citep{SunSchubert1993,CHRISTENSEN2002,Simitev2003,Simitev2005b,GILLET2007}
show that immediately after onset convection assumes the form of
shape-preserving columns parallel to the rotation axis and drifting in
azimuthal direction with time-independent azimuthally averaged properties. From a frame of
reference drifting with the convection columns the entire
pattern appears steady. Differential rotation is generated through the
action of Reynolds stress caused by the spiralling cross section of
the columns. Further away from the onset bifurcations occur that 
break all available temporal and  spatial symmetries of the problem
and lead to flows with increasingly chaotic temporal and
spacial structures and dependence. 
{These transitions are captured well by tracking kinetic energy
components e.g.~\citep{Simitev2003,Simitev2015}. 
Figure~\ref{fig010} shows time-averaged values of the kinetic energy
density components of the flows along the selected ``slices''
of the  $\RCi-\RTi$ plane and demonstrates that a similar approach is
also useful in tracking double-diffusive regimes.}

Increasing the value of $\RTi$ in  panels
\ref{fig010}(a) and (c) one observes the regimes of rotating fingering
convection D, compositionally dominated overturning convection C 
and thermally dominated  overturning convection B and the transitions
between them. In figure  \ref{fig010}(c), regime D appears as an 
isolated island of instability separated from regime C by a
quiescent region. The amplitude of fingering convection motions
attains a maximum at a value of $\RTi$ situated within the interior of
the island of instability and decreases towards its ends. 
At a somewhat larger value of $\RCi$ depicted in figure
\ref{fig010}(a), the domains of regimes D and C join up but the
transition between them is well visible in the pronounced 
dip in kinetic energy components. An abrupt transition from regime C
to regime B occurs at about $\RTi=2.7\times10^5$ in both panels (a)
and (c) of figure \ref{fig010}. In figure \ref{fig010}(e), the
transition between the regimes of rotating semi-convection A and
thermally-dominated overturning convection B is shown. This transition
is also abrupt and occurs at about $\RTi=2.8\times10^5$. The
transition value $\RTi=2.8\times10^5$ appears to be situated along the
continuation of the thermally-dominated overturning convection branch
of the critical curve for the onset. 
Panels \ref{fig010}(b,d,f) show kinetic energy densities as functions
of the compositional Rayleigh number $\RCi$ for several selected values of
the thermal Rayleigh number $\RTi$. Panel (b) shows the
energy density components of cases in regime B. A transition to a branch
with dominant differential rotation is observed at about
$\RCi=9\times10^5$. This is a transition internal to regime B, namely a
transition to chaotic convection. In figure \ref{fig010}(d), regimes A 
and C are observed separated by a wide quiescent region of no
convection. A gradual increase of the amplitude of the flows is
seen as convection is driven away from the critical curve of linear
onset. Finally, figure \ref{fig010}(f) shows the energy density 
components of flows in regime D as they are driven away from the
onset neutral curve.

%[RATIOS OF ENERGY COMPONENTS]
% $\Eft$ and $\Efp$ 
The fluctuating toroidal energy $\Eft$ and the fluctuating poloidal
energy $\Efp$  are kinetic energy components corresponding to flows in
the azimuthal direction and flows within meridional planes, respectively.  
These two components show very similar behaviour in all four
convective regimes as seen in figure \ref{fig010} because locally any flow
pattern consists of both azimuthal and radial motions. The fluctuating
kinetic energies dominate the flows in the regimes of rotating
semiconvection A, fingering convection D, and compositionally dominated
overturning convection C as well as in the vicinity of the onset
transition of the regime of thermally dominated overturning convection
B.  It may be observed in figure \ref{fig010} that in all cases the
approximate relation $\langle\Efp \rangle_t\approx 0.3
\langle\Eft\rangle_t$ is satisfied regardless of the particular
convection regime. 
% $\Emp$ 
The mean poloidal energy $\langle\Emp\rangle_t$ is the energy
contained in the mean meridional circulation. Since the effects of
rotation strongly suppress the motions in the axial direction this
component remains small, and this is seen on all convection branches
in figure \ref{fig010}.
% $\Emt$
The mean toroidal energy $\langle\Emt\rangle_t$ is the kinetic energy
contained in differential rotation.  
On all branches shown in figure \ref{fig010} except in the regime of
thermally-dominated overturning convection B, the time-averaged value of
$\langle\Emt\rangle_t$  is small and comparable to $\langle\Emp\rangle_t$ . There are a number of reasons
for this. Firstly, compositional regimes A and D appear to be limited
in range and transitions to the overturning regimes B and C occur
before their energy components, including $\langle\Emt\rangle_t$, can grow
significantly. In addition, the two regimes A and D show broad convection
structures that are not prone to spiralling, see further below. The regime 
of compositionally-dominated overturning convection C is
characterised with a large value of the Schmidt number $\Sc$ and
behaves much like purely thermal convection at large values of $\Pr$ where
spiralling and consequently differential rotation is known to be weak
\citep{Simitev2005a}. Differential rotation starts to grow only when
the regime of thermally-dominated overturning convection B is entered.
Since the toroidal fluid motions do not have a radial component, it 
is the poloidal motions which are directly associated with the
transport of heat and material between the boundaries. The
differential rotation, however, helps dynamo generation by the
$\Omega$-effect.

%[TIME DEPENDENCE]
Regarding time-dependence, compositional regimes A and D are
time-independent since transitions to regimes B and C occur before
their near-onset patterns can become unstable and develop
chaotic structures. Regimes B and C are time dependent and exhibit the
usual sequence from steady state via time-periodic oscillations to
chaotic behaviour \citep{Simitev2003}.
\begin{figure*}
        \centering
\includegraphics[width=\textwidth,clip]{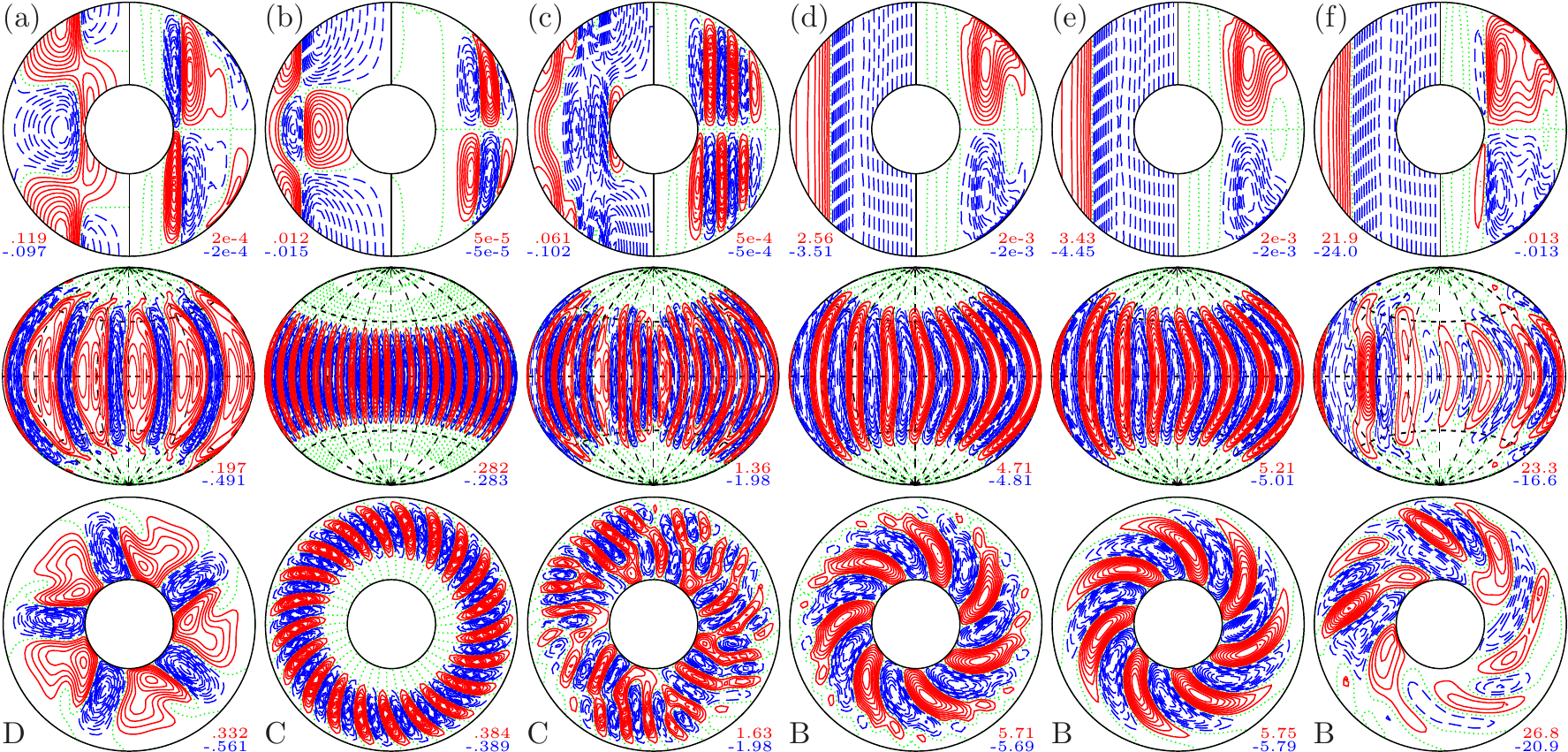}
        \caption{Flow components in the case
$\eta=0.35$, $\tau=10^4$, $\Pr=1$,  $\Sc=25$, 
$\RCi=4.5\times10^5$, and increasing values of $\RTi=-2\times$,
$1\times$, $2\times$, $2.8\times$, $3\times$, $5\times10^5$, from (a) to (f),
respectively, (all part of the sequence shown in figure 
\ref{fig010}(c)).
The first plot in each column shows isocontours of
azimuthally-averaged $\overline{u}_\varphi$ (left half) and
streamlines $r \sin \theta (\partial_\theta \overline{v}) = $
const. (right half) in the meridional plane. The second
plot shows isocontours of $u_r$ at $r = r_i + 0.5$ mapped to the
spherical surface using an isotropic Aitoff projection. The third plot
shows isocontours of $r \partial_\varphi v = $ const. in the
equatorial plane.
{The minimal and the maximal values of each field are listed under the
corresponding plot.} The isocontours are equidistant with positive isocontours shown by
solid lines, negative isocontours shown by broken lines and the zeroth
isocontour shown by a dotted line in each plot. 
All contour plots are snapshots at a fixed representative moment in time. Labels D, C and B
denote the dominant regime of convection. 
(Colour online)
}
\label{fig4}
\end{figure*}

\subsection{{Typical flow patterns and interactions}}
\begin{figure*}
        \centering
\includegraphics[width=0.84\textwidth,clip]{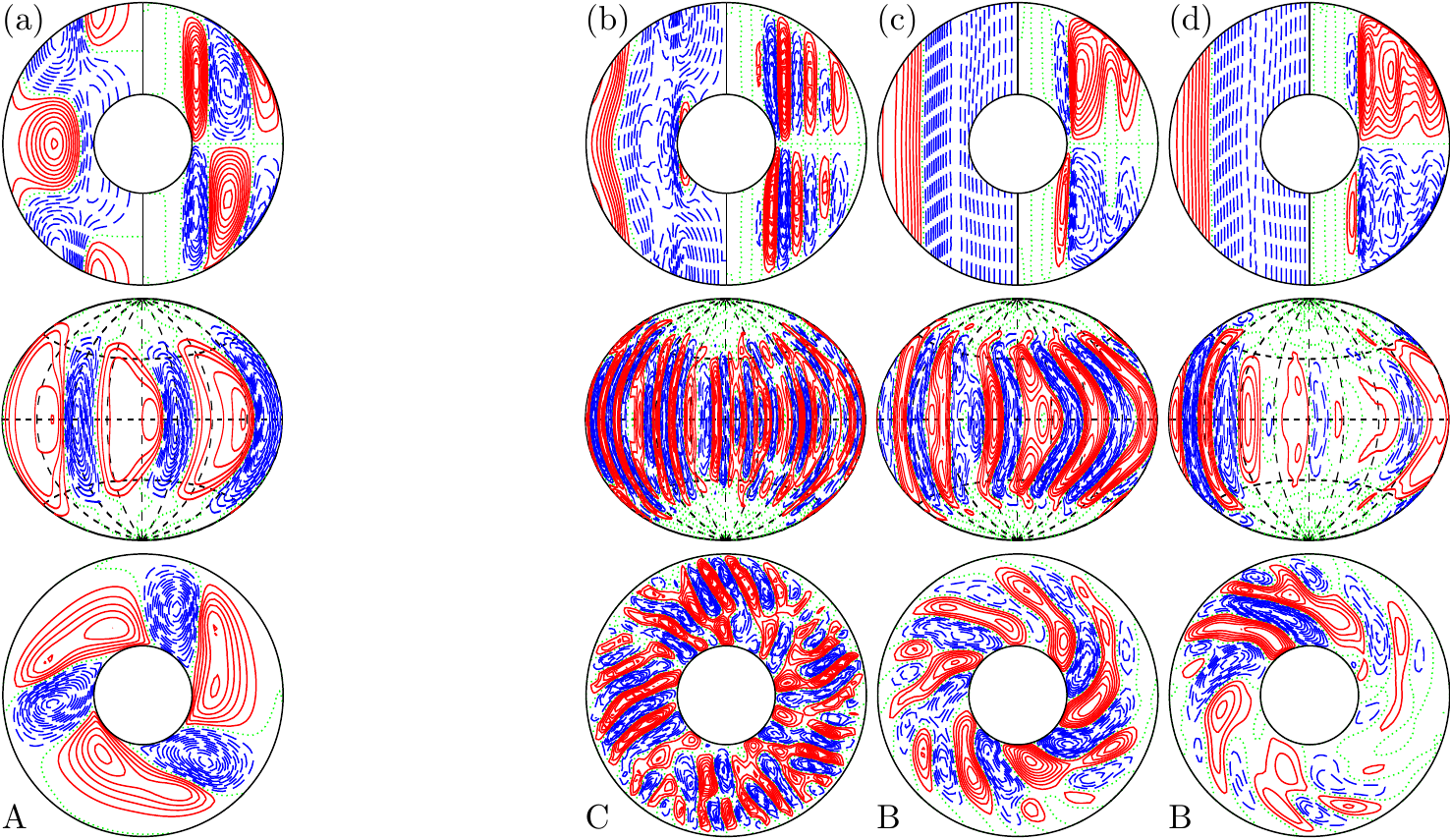}
        \caption{Flow components in the case
$\eta=0.35$, $\tau=10^4$, $\Pr=1$,  $\Sc=25$, and values
of the  Rayleigh numbers as follows (a) $\RCi=-4\times10^5$ and
$\RTi=2\times10^5$, (b) $\RCi=8\times10^5$ and $\RTi=2\times10^5$, (c) $\RCi=8\times10^5$
and $\RTi=4\times10^5$ and (d) $\RCi=8\times10^5$ and $\RTi=6\times10^5$.  The same
flow components are shown as in  figure \ref{fig4}. Labels A, C and B
denote the dominant regime of convection.
(Colour online)
}
        \label{fig5}
\end{figure*}

%[FLOW PATTERNS]
Examples of typical flow patterns across the four convective regimes
are presented in figures \ref{fig4} and \ref{fig5}. 
Figure \ref{fig4}, in
particular, includes selected cases located along the slice of
the $\RCi-\RTi$ plane shown in figure \ref{fig010}(c). The first
column of \ref{fig4} shows a typical case in the state of rotating
fingering convection D. At a value $\RCi=-2\times10^5$ this case is
located nearly in the middle of the regime region. The equatorial
cross section of the poloidal streamlines, shown in the lowermost plot
of figure \ref{fig4}(a), reveals a pattern of wave number 5 consisting
of pairs of a narrow column with anticlockwise flow and a 
broad column with clockwise flow. The broad column appears as composed
of two structures with clockwise flow adjacent to each other
and with the one located closer to the inner core being more
vigorous. The convective columns and their vertical morphology are
visible in the plot of the radial velocity on a spherical surface
located at the middle of the shell, shown in the middle row of
figure \ref{fig4}(a). The plot of the radial velocity also reveals
that there is little or no polar convection. Differential rotation and 
azimuthally averaged meridional circulation are plotted in the left
and the right halves, respectively, of figure \ref{fig4}(a).
The differential rotation profile is rather peculiar. Differential
rotation is symmetric with respect to the equatorial plane and shows a
prominent tube of retrograde flow situated at the equatorial region
and extending to about $30^\circ$. On top of this tube there are
two azimuthal jets of similarly strong  prograde  differential
rotation. The shear layer between the jet of retrograde and the jets
of prograde rotation is almost horizontal and the two prograde jets
link to each other via a thin layer going behind  the equatorial
retrograde jet. Two weak retrograde jets are finally visible in the
polar regions. The meridional circulation is antisymmetric with
respect to the equatorial plane {and consists of} three cells oriented
parallel to the axis of rotation.
\citep{Simitev2003,Simitev2005b}. 

At $\RTi=10^5$ figure \ref{fig4}(b) illustrates a
compositionally-dominated overturning convection case in regime C as
expected from linear analysis. The case is in the vicinity of the
linear onset and the convection pattern changes dramatically from the
case just described. The poloidal streamlines plotted in the
equatorial plane exhibit a dominant azimuthal wave number of 21. The
convective columns are rather thin and adjacent to the outer spherical
surface rather than to the inner core. There is no polar convection as
rotation inhibits polar motions. Meridional circulation is
antisymmetric withe respect to the equator and consists of three
cells. The convection columns are spiralling very weakly and thus only
weak differential rotation is generated. The differential rotation is
now prograde  at the outer spherical surface and at the equatorial
region. Pronounced retrograte jets appear at the poles while a
pronounced prograde jet appears at the inner surface near the
equatorial region. 

At  $\RTi=2\times10^5$ figure \ref{fig4}(c) shows a more
chaotic compositionally-dominated overturning convection case in regime
C. The equatorial streamlines show convection columns with a wave
number of 20 near the outer surface which now extend all the way
to the inner core where clockwise columns coalesce to a wavenumber of
about 10. The columns extend to higher latitude as seen in the plot of
the radial velocity on the spherical surface. The mean meridional
circulation is now five-cell and the differential rotation has become
stronger with the two large polar jets extending towards the equator
and coalescing. 

The last three columns of figure \ref{fig4} exhibit three cases 
of thermally-dominated convection in regime B that become increasingly 
chaotic with the increase of the value of $\RTi$. The poloidal
streamlines in the equatorial plane  reveal a flow pattern of
7 pairs of convection columns that are anchored at the inner core and
strongly spiral outwards. The mean meridional streamlines form a large
one-cell antisymmetric pattern and the differential rotation is
exhibits a strictly geostrophic pattern constant on cylinders parallel
to the tangent cylinder. The case with $\RTi=2.8\times10^5$ shown in
figure \ref{fig4}(d) is situated very
nearly at the transition from state C which is evidenced by a weak
modulation of the convective column tips near the outer spherical
surface. The case with $\RTi=5\times10^5$ shown in figure
\ref{fig4}(f) is in a chaotic state known as a
localized convection where differential rotation is so strong that it
shears-off the convective columns so that convection is weak or
suppressed in approximately half the shell volume. These properties
are essentially identical to the properties of purely thermal
convection at comparable parameter values

The last three columns of figure \ref{fig5} also illustrate the
transition from compositionally- to thermally-dominated convection,
i.e. from regime C to regime B, but in a more strongly chaotic cases
with a larger value of $\RCi=8\times10^5$. It is interesting that the
C to B transition and the corresponding patterns of convection are
very similar to the ones just described despite the significant
increase in compositional driving.  

Finally at $\RCi=-4\times10^5$, the first column of figure \ref{fig5}
illustrates a typical rotating semi-convection case in regime A. An
azimuthal wave-number 3 is exhibited by the poloidal streamlines in
the equatorial plane. The asymmetry between narrow columns with
anticlockwise flow and broad columns with clockwise flow featured in
by fingering regime D is also present here. The profiles of mean
meridional flow and of the differential rotation are also very similar
to the corresponding patterns of the fingering regime but are in
reversed direction, i.e. a similar in shape equatorial jet of
differential rotation is seen however in prograde rather than in
retrograde direction. 

In this section cases with small to moderate driving were used in
order to clearly illustrate the patterns of convection and the
transitions between the four regimes of rotation double-diffusive
convection. More strongly driven convection cases are presented in the
next section  where a sufficient intensity of the flow is a
prerequisite for dynamo action. 

\section{Dynamos driven by thermo-compositional convection}

{Having determined how the four thermo-compositional regimes
evolve at finite-amplitudes, in this section we describe  attempts (a)
to find dynamo 
solutions in the semi-convection and fingering regimes and
(b) to assess differences between dynamos generated by
thermally-dominated and chemically-dominated overturning convection.}
\begin{figure*}
        \centering
\includegraphics[width=\textwidth,clip]{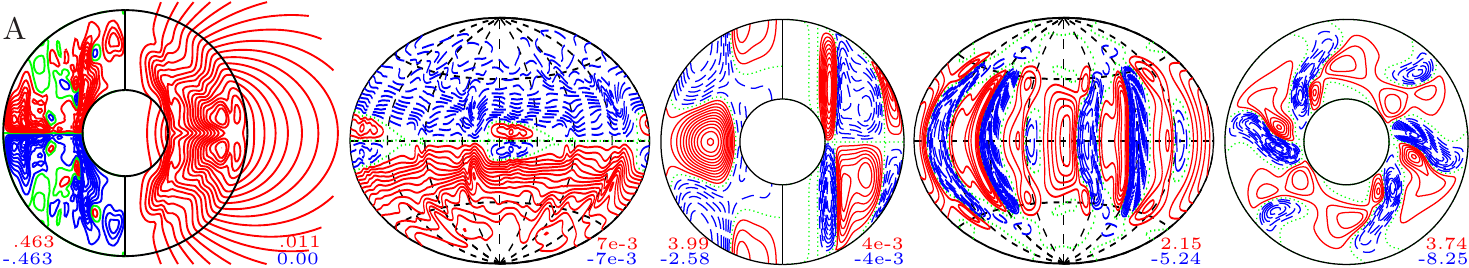}
\caption{Flow and field structures of a dynamo in convective regime A
  at parameter values $\eta=0.35$, $\tau=10^4$,
  $\Pr=1$,  $\Sc=25$, $\RTi=2\times10^5$, $\RCi=-6\times10^6$, 
and $\Pm=300$. 
The first plot shows isocontours of $\overline{B_\phi}$ (left
half) and meridional fieldlines $r \sin \theta \partial_\theta \overline{h} =$ const. (right half). 
The second plot shows isocontours of
radial magnetic field $B_r$ at $r = r_o + 0.1$ in isotropic Aitoff
projection.  
The last three plots show the same flow components as in  figure \ref{fig4}.
{The minimal and the maximal values of each field are listed under
  the
corresponding plot. The isocontours are equidistant with positive
isocontours shown by
solid lines, negative isocontours shown by broken lines and the zeroth
isocontour shown by a dotted line in each plot.}
All contour plots are snapshots at a fixed representative moment in
time
(Colour online).}
\label{fig5dynA}
\end{figure*}

\subsection{{Dynamo action in the semi-convective and fingering regimes}}
%\red{[\%DESCRIPTION OF A DYNAMO IN STATE A]}
% /home/staff2/dynamo/dat/00_Results/2019-06-19Energies/selecteddynamos/cp1t10ri200000s25rci-6000000m2p300hh.eb
A typical dynamo generated by rotating semi-convective flows in regime A at
$\RTi=2\times10^5$ and $\RCi=-6\times10^6$ and a value of the magnetic Prandtl
number $\Pm=300$ is presented in figure \ref{fig5dynA}.
This dynamo exhibits a chaotic time dependence, {as shown in the
lowermost plot of figure \ref{fig08}(a),} even though snapshots
at different times remain rather similar to the one shown in  figure
\ref{fig5dynA}. The ratio of toroidal to poloidal magnetic energy is 
$E_\text{tor}^\text{magn}/E_\text{pol}^\text{magn} = 9.98$ so that
much of the magnetic field is confined to the core. The ratio of
$E_\text{dip}^\text{magn}/E_\text{quadr}^\text{magn}=3272.9$ 
so that the dynamo is almost perfectly dipolar. However, the 
ratio of total magnetic to total kinetic energy is rather small at
$E_\text{magn}/E_\text{kin} = 0.0042$.  
Because the magnetic field is so weak compared to the intensity of
the flow, convection is nearly unaffected by dynamo action. Indeed, the
spatial patterns of the flow remain very similar to the ones
described in relation to the corresponding case shown in figure
\ref{fig5}(a). The most pronounced difference is that {the
dominant azimuthal wavenumber of the flow is $m=6$, as seen in the $m$-spectrum of
the kinetic energy in figure \ref{fig08}(a), and} is larger that the
azimuthal wave number in the non-magnetic case shown in the first
column of \ref{fig5}(a) which is likely due to the smaller value of the
compositional Rayleigh number $\RCi$ used here. A plot of the azimuthally
averaged toroidal fieldlines in the meridional plane are shown in the
left half of the first plot of figure \ref{fig5dynA}. The plot reveals a pair of
toroidal flux tubes of opposite polarity located near the outer core
above and below the equator. A second pair of elongated toroidal flux
parallel to the axis of rotation are situated at the cylinder tangent
to the inner core. This profile of the azimuthally averaged toroidal
fieldlines correlates well with the profile of the differential
rotation shown in the left half of the third plot in \ref{fig5dynA}
which exhibits regions of shear near the equator and on the tangent
cylinder.
The azimuthally averaged poloidal  fieldlines in the meridional plane
are shown in the right half of the first plot of figure \ref{fig5dynA} and exhibit a
typical dipolar structure {as also evident in by the $l$-spectrum
of magnetic energy shown in figure \ref{fig08}(a)}. Besides the apparent large-scale dipolar
field, the plot of the radial magnetic field at surface exhibits two
small patches of inverted polarity near the equator. 
{Because the ratio of total magnetic to total kinetic energy is very
small while at the same time the value of the magnetic Prandtl number
$\Pm=300$ needed to support dynamo action in this case is very large
it is unlikely that dynamos in state A are of relevance in planetary
and stellar magnetism.}
\begin{figure*}
        \centering
\includegraphics[width=\textwidth,clip]{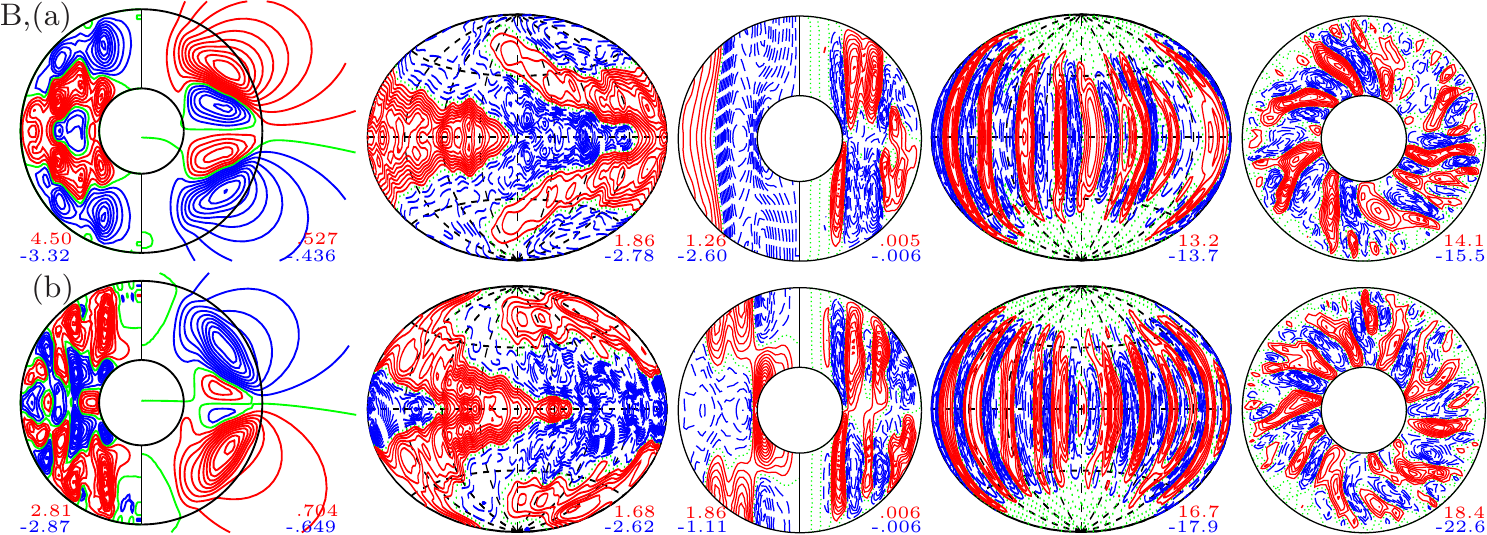}
\caption{Oscillation of a quadrupolar dynamo in
          convective regime B at parameter values $\eta=0.35$,
          $\tau=10^4$, $\Pr=1$,  $\Sc=25$, $\RTi=3\times10^5$,
          $\RCi=1.2\times10^6$,  and $\Pm=15$. Snapshots are
          shown (a) at the beginning and (b) at the end of
          a half a period the oscillation with time lapse $\Delta t=5.0$ 
The same flow and field components are shown as in figure
\ref{fig5dynA}.    
(Colour online)}
\label{fig7dynB}
\end{figure*}

We have been unsuccessful in our efforts to obtain a non-decaying
dynamo in regime D of rotating fingering double-diffusive convection.
{The values of both $\Pm$ and $\RCi$ were increased along the line
  $\RTi=-200000$, but with the computational constraints of the code
only decaying dynamos solutions were obtained.}

\subsection{{Dynamo action in the overturning regimes}}
It is off interest to investigate how the sharp transition between
regimes  B to C affects dynamo behaviour and whether the
properties of the magnetic field also exhibit an abrupt transition.
An example of a dynamo in the regime of thermally-dominated overturning
convection B is presented in figure \ref{fig7dynB}
at values of the magnetic Prandtl number $\Pm=15$, the
compositional Rayleigh number $\RCi=1.2\times10^6$ and the thermal Rayleigh
number $\RTi=3\times10^5$, respectively.
This is a strong field dynamo with a ratio of total magnetic to 
kinetic energy $E_\text{magn}/E_\text{kin} = 1.01$.  
The predominant dynamo morphology is quadrupolar as 
evidenced by the {the $l$-spectrum of magnetic energy shown in
figure \ref{fig08}(b)} and by the ratio of dipolar to quadrupolar energy components
$E_\text{dip}^\text{magn}/E_\text{quadr}^\text{magn}=0.02$. 
The component detectable outside of the dynamo generating region is
strong as measured by the ratio of toroidal to poloidal magnetic
energy  $E_\text{tor}^\text{magn}/E_\text{pol}^\text{magn} = 1.22$. 
As is typical for quadrupolar dynamos this example exhibits regular
periodic oscillations in time.  
The oscillations are well visible in the plots of all magnetic field 
components and represent a dynamo wave propagating from the equator to
the poles.
Emergence of magnetic flux is initiated at the equator near the inner
core boundary as can be best seen in the plots of the azimuthally
averaged toroidal and poloidal fieldlines in the first column of figure \ref{fig7dynB}. These new flux tubes proceed to
grow and drift to higher latitudes pushing old flux tubes upwards in
the northern hemisphere and downwards in the southern hemisphere. At
high latitudes the magnetic flux tubes in question become weaker
gradually dissipate and are replaced in a similar fashion by new flues
of the opposite polarity.  The dynamo wave exhibits a phase shift in
azimuthal direction in the sense that at any given moment the
pole-ward drifting flux tubes appear at various latitude for various
azimuthal angles thus forming a pair of V-shaped spiraling structures
as seen in the contour plot of the radial magnetic field $\B_r$.
The frequency of oscillations is proportional to the square roots of
the magnetic Prandtl number and the kinetic helicty and to the fourth
root of the differential rotation \citep{Simitev2006b}.
Although this is a strong field dynamo, flow structures are weakly
affected by the magnetic field. {The dominant azimuthal wavenumber
of the flow is $m=8$ as seen in the $m$-spectrum of
the kinetic energy in figure \ref{fig08}(b) and other flow features} remain much as described in
relation to figures \ref{fig4} and \ref{fig5} above. An exception to
this is the strong effect of the magnetic field on differential
rotation which is significantly reduced in amplitude and oscillates
between a geostrophic and conical profile as seen in the left half of
the third column of figure \ref{fig7dynB}. The weaker differential
rotation is as a result less able to shear-off the azimuthally
drifting convective columns. This example is very similar to
oscillating quadrupolar dynamos familiar from purely-thermal models
e.g. \citep{Simitev2005b}. It is perhaps surprising that a quadrupolar
dynamo occurs at a rather larger value of the magnetic Prandtl number
than in the former case, e.g. see figure 3 of \citep{Simitev2005b}. 
\begin{figure*}
        \centering
\includegraphics[width=\textwidth,clip]{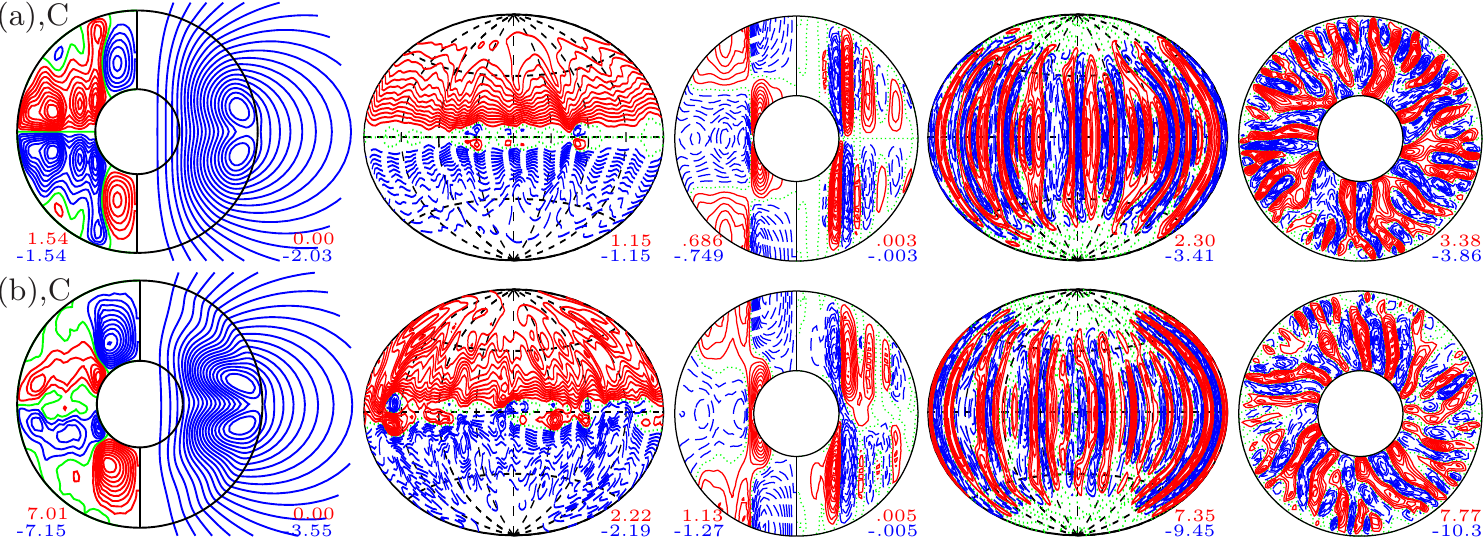}
        \caption{
Two typical dynamos in convective regime C at parameter values 
$\eta=0.35$, $\tau=10^4$, $\Pr=1$,  $\Sc=25$,  and 
(a)  $\RTi=-2\times10^5$, $\RCi=1.2\times10^6$ and $\Pm=30$ and  (b) $\RTi=2\times10^5$,
$\RCi=1.2\times10^5$ and $\Pm=30$.
The same flow and field components are shown as in figure
\ref{fig5dynA}.
(Colour online)
}
\label{fig6dynC}
\end{figure*}

%\red{[\%DESCRIPTION OF DYNAMOS IN STATE C]}
Two typical dynamos in regime C of rotating compositionally-dominated
overturning convection are illustrated in figure \ref{fig6dynC} at a
value of the magnetic Prandtl number $\Pm=30$, a value of the
compositional Rayleigh number $\RCi=1.2\times10^6$ and thermal Rayleigh
numbers $\RTi=-2\times10^5$ and $\RTi=2\times10^5$, respectively.
Both dynamos appear very similar, indeed. They are strong field
dynamos with a ratio of total magnetic to total kinetic energy $E_\text{magn}/E_\text{kin} = 3.31$ and $2.61$ the dynamos with
$\RTi=-2\times10^5$ and $\RTi=2\times10^5$, respectively.
The predominant dynamo morphology is dipolar as evidenced also by the
ratio of dipolar to quadrupolar energy components which is 
$E_\text{dip}^\text{magn}/E_\text{quadr}^\text{magn}=28.2$, and
$8.87$ for the first and the second dynamos, respectively. {The
dipolar structure of the second dynamo is confirmed by the
$l$-spectrum of the magnetic energy in figure \ref{fig08}(c).} 
The externally observable components the dynamos are weak as
measured by the ratio of toroidal to poloidal magnetic energy and are
$E_\text{tor}^\text{magn}/E_\text{pol}^\text{magn} = 0.32$ and $0.56$,
for the first and the second dynamos, respectively.
As is typical for dynamos dominated by dipolar components both dynamos
are non-oscillatory and, while the time series of their kinetic and
magnetic energy components are chaotic, the spacial patterns of the
velocity and magnetic fields remain very similar to the ones shown in
figure \ref{fig6dynC}.
The azimuthally averaged toroidal fieldlines shown in
the left halves of the first column of figure \ref{fig6dynC} exhibit
magnetic flux tubes antisymmetric with respect to the equator. In
particular, {a} pair of strong flux tubes appear in the polar regions
adjacent to the axes of rotation. The case with $\RTi=-2\times10^5$ shows an
equally strong pair of ``butterfly'' shaped flux tubes at mid to low
latitudes. A similar pair appears in the case $\RTi=2\times10^5$ but seems
to be smaller and weaker in comparison to the polar flux tubes of this
case. The plots of the radial magnetic field also exhibit similar
equatorially antisymmetric patterns when projected onto the spherical
surface of the shell. Although both cases are strong field dynamos, their flow structures
are not really affected by the magnetic field and remain much as described
in relation to figures \ref{fig4} and \ref{fig5} above. Again an
exception to this is the profound effect of the magnetic field on
differential rotation which being weaker than the one of the cases in
regime B, is further reduced in amplitude and is modified to assume a
profile featuring a retrograde jet in the equatorial region.
\begin{figure*}
\centering
\includegraphics[width=\textwidth,clip]{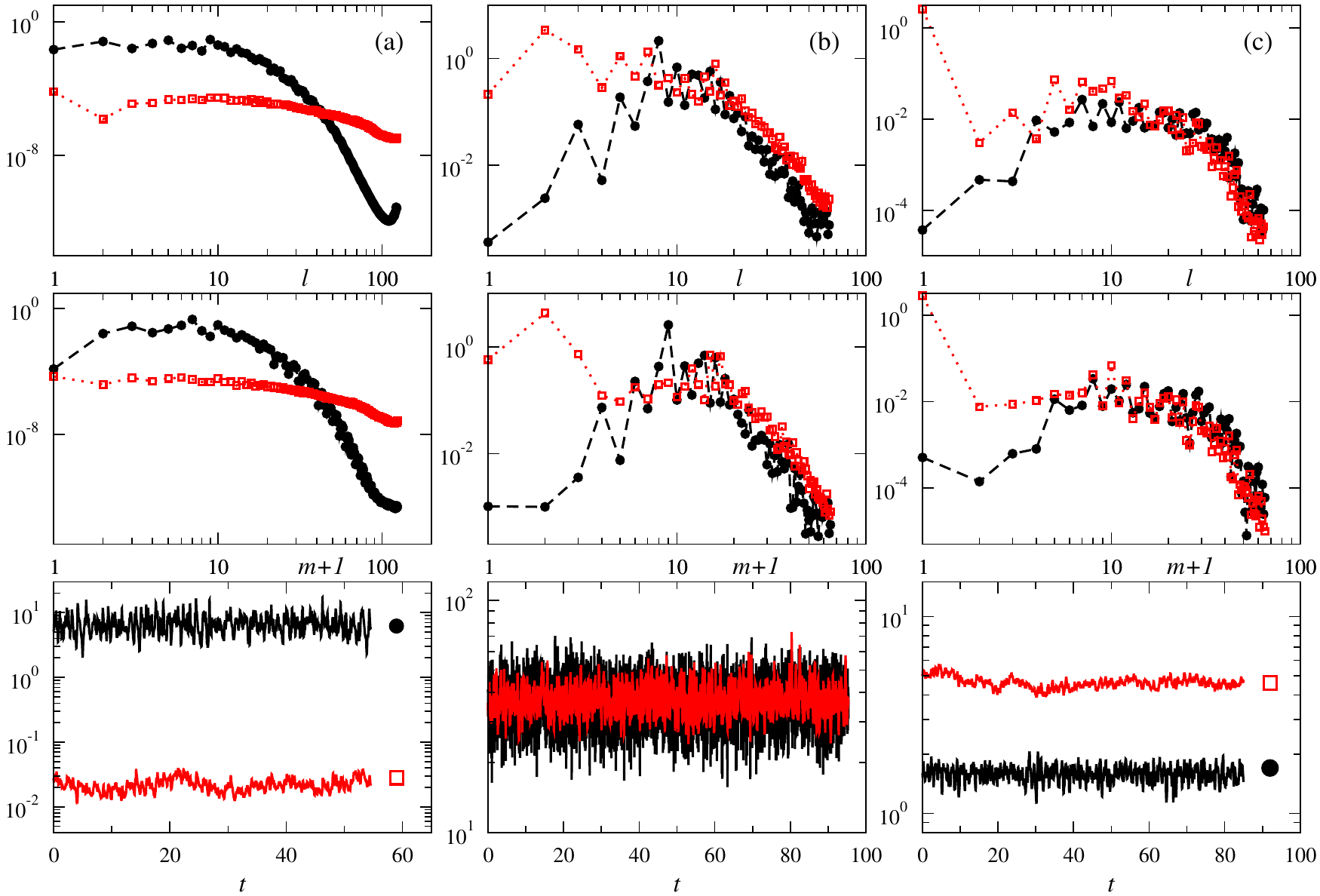}
\caption{{
Instantaneous radial averages of kinetic energy spectra (black solid circles)
and magnetic energy spectra (red empty squares) as functions of
the spherical degree $l$ (top row) and azimuthal order $m+1$
(middle row). The bottom row shows total kinetic energy densities (black line marked
with black solid circle) and magnetic energy densities (red line marked
with red empty square) as functions of time for the dynamos shown in
(a) figure 
\ref{fig5dynA}, (b) figure \ref{fig7dynB} and (c) figure \ref{fig6dynC}(a).
(Color online).}}
\label{fig08}
\end{figure*}

It is interesting to note that an abrupt transition  in magnetic
field properties seem to exist when the boundary of the convective
regimes B and C is traversed.  Indeed, the relatively small change in
the value of the thermal Rayleigh number $\RTi$ {from $3\times10^5$ to
$2\times10^5$} needed to transition between the dynamos shown in
figures \ref{fig7dynB} and \ref{fig6dynC}{(b)} produces magnetic fields
of very different intensity and morphology. In addition, there is a
pronounced difference in the value of the magnetic Prandtl number
required to obtain non-decaying dynamos in the two regimes in
question. {Indeed, $\Pm=15$ for the example shown in figure \ref{fig7dynB}
  while $\Pm=30$ for the  example shown in figure \ref{fig6dynC}(b),
  with both solutions being near the onset of dynamo action.}

\section{Conclusion}
{
The main goals of this work are to establish the possibility of dynamo
action arising due to double-diffusive flows on stably stratified
background and to assess how dynamos generated by thermally-and
chemically-dominated overturning convection differ from each other. 
}
Guided by previous linear stability analysis of the onset of
convection in this configuration \citep{Silva2019}, we
traced out in the nonlinear domain the four distinct
regimes of rotating double-diffusive convection, namely rotating
semi-convection, rotating thermally-dominated overturning convection,
rotating compositionally-dominated overturning convection and rotating 
fingering convection.  

% Model and why
Since the configuration space of the problem is now significantly
larger that that of purely thermal
convection, we restrict the attention to a set of carefully 
selected parameter values and model assumptions that allow for (a) a 
systematic access to all four flow regimes and at the same
time (b) for direct comparison of results to purely thermal
simulations readily available in the literature 
e.g.~\citep{Simitev2003a,Simitev2005b,Simitev2006b}. To this end, 
internally distributed heat and concentration sources, stress-free
velocity and fixed temperature and concentration boundary conditions
were used at the electrically insulating inner and outer spherical
surface following the articles cited above. Both thermal and 
compositional Rayleigh numbers $\RTi$ and $\RCi$ were varied while all
other parameters were fixed at relatively modest but computationally
feasible values $\Pr=1$, $\Sc=25$, $\tau=10^4$, $\eta=0.35$, and the
magnetic Prandtl number was selected so as to be just over the onset
of dynamo action. 

% Provide answers to the following open questions posed in the intro
% and studied in the paper
We find that convection instability pockets identified in the linear
stability analysis \citep{Simitev2011,Silva2019} survive nonlinear
interactions and that the boundaries of the four convective regimes
can be traced towards finite-amplitudes in the configuration space for,
at least, several times the critical values of $\RTi$ and $\RCi$. We
confirm that boundaries remain sharp resulting in abrupt transitions
between convective regimes as noted by \citet{Breuer2010} for the boundary
between the thermally-dominated and the compositionally-dominated
overturning convection at the parameter values they used. Typical
velocities of flows in the semi-convecting and the fingering regimes
remain relatively small compared to those in the two overturning
regimes, and zonal flows remain weak in all regimes apart from the
regime of thermally-dominated overturning convection. Finite-amplitude
compositionally-dominated overturning convection exhibits
significantly narrower azimuthal structures compared to all other
regimes while the convective columns of the semi-convecting and the
fingering regimes are particularly wide but with some notable asymmetry
between clockwise and counter-clockwise vortices with the latter being
narrower. The thermally-dominated overturning regime retain properties
very similar to that of purely thermal convection, with differential
rotation becoming the dominant flow component as driving is increased.

%Dynamos
We find that dynamo action occurs in all regimes apart from the regime
of fingering convection. In the latter increasing the value of the magnetic
Prandtl number and venturing, as far as feasible, further out to more
strongly driven flows failed to sustain the initial magnetic field
seeds. 
Dynamo action persists in the semi-convective regime but it is very
much impaired by the small intensity of the flow and the by very weak
differential rotation which makes conversion of poloidal to toroidal
field problematic. Unrealistically, large  values of the magnetic
Prandtl number were required to obtain a non-decaying dynamo solution
in this case. {The results of \cite{Silva2019}, suggest that the
region of semiconvection becomes wider as rotation is
increased, and since stronger rotation also promotes stronger mean
zonal flows we speculate that dynamos driven by semiconvection may be
found more easily and at larger magnetic diffusivity at higher values
of the Coriolis parameter $\tau$.}
Both regimes of overturning convection easily support magnetic field
generation. The dynamos in the thermally-dominated regime 
include oscillating dipolar, quadrupolar and multipolar cases similar
to the ones known from our earlier parameter studies at comparable
parameter values
e.g. \citep{Simitev2003a,Simitev2005b,Simitev2006b,Simitev2011b}.
{For this reason, comparison with scaling laws for dynamos driven
by thermal convection e.g.~\citep{CHRIST&AUBERT2006} is not expected to
yield additional insight.}
Due to the significantly weaker zonal flow dynamos in the
compositionally-dominated regime show much more subdued temporal
variation and remain predominantly dipolar as also reported by
\citet{Takahashi2014}.  

{To meet the goals of this study, it was necessary to use parameter
values removed from both geo- and planetary estimates.} To rectify
this, it is desirable to approach more ambitious values allowed by
modern computing resources.  
In the configuration studied, we did not find a regime where stable quiescent
layers or layers with a distinct flow  properties form spontaneously
below the core-mantle boundary and coexists with flow in the bulk. It
is of significant interest
\citep{Jones2015,Olson2018,Wicht2019,Bouffard2019} to investigate
whether such layers may form under different boundary conditions for
the concentration and the temperature field and or different internal
source-sink distributions. These issues remain interesting avenues for
future work.

\section*{Acknowledgements}
This work was supported by the Leverhulme Trust [grant number
RPG-2012-600].

\markboth{\rm J.F.~MATHER AND R.D.~SIMITEV}{\rm GEOPHYSICAL AND ASTROPHYSICAL FLUID
DYNAMICS}

%\bibliographystyle{gGAF} 
%\bibliography{referencesDynamosGGAF}

\end{document}